\documentclass{mem} 

\usepackage{natbib,graphicx,amssymb,txfonts,balance}
\idline{75}{282}

\begin{document}

\title{Confining the high-energy cosmic rays}
\author{K.M.~Schure\inst{1}
\and A.R.~Bell\inst{1}}
\institute{Department of Physics, University of Oxford, Clarendon Laboratory, Parks Road, Oxford OX1 3PU, United Kingdom
\email{k.schure1@physics.ox.ac.uk}
}

\abstract{
Diffusive shock acceleration is the prime candidate for efficient acceleration of cosmic rays. Galactic cosmic rays are believed to originate predominantly from this process in supernova remnant shock waves. Confinement of the cosmic rays in the shock region is key in making the mechanism effective. It has been known that on small scales (smaller than the typical gyroradius) high-amplitude non-resonant instabilities arise due to cosmic ray streaming ahead of the shock. For the efficiency of scattering of the highest energy cosmic rays it is of interest to determine the type of instabilities that act on longer length scales, i.e. larger than the cosmic ray gyroradius. We will present the results of our analysis of an instability that acts in this regime and will discuss its driving mechanism and typical growth times.
}
\maketitle{}

\section{Introduction}
\label{sec:intro}
Galactic cosmic rays are widely believed to be accelerated, be it predominantly or not, in supernova remnants (SNRs). If solely SNRs are responsible for the acceleration, a significant fraction of kinetic energy has to be transferred onto cosmic rays. Additionally, the acceleration needs to be efficient enough to arrive at energies typical of the knee in the spectrum, being around $10^{15}$~eV. In order to get those acceleration rates, Bohm diffusion is normally assumed, since this yields the quickest isotropisation needed for the confinement of cosmic rays around the shock region. In reality, the diffusion is dependent on the power in magnetic fluctuations on the scale of the gyroradius. Scattering of the highest-energy cosmic rays requires a significant component of magnetic field fluctuations on the order of parsec scales. A cosmic-ray streaming instability that initialises growth on those scales is derived from the Boltzmann equation for the cosmic rays, coupled to the MHD equations of the background plasma. Typical growth times are presented.

\section{Cosmic ray streaming instability in the long-wavelength limit}

It has long been known that cosmic rays can act to trigger magnetic fluctuations. The first calculations on this were in the resonant regime, where energy is transferred from the cosmic rays to magnetic fluctuations that have the same scale as the gyroradius of the cosmic rays \citep{1967Lerche, 1969KulsrudPearce, 1974Wentzel, 1975Skilling}. When the perturbed magnetic field grows to values similar to the background field, the resonance condition is lost, and any magnetic field amplification beyond $\delta B/B \approx 1$ must be non-resonant. 

With the development of diffusive shock acceleration, the need for efficient scattering around the shock region arised in order to get to cosmic ray energies of $10^{15}$~eV. Additionally, observational evidence of narrow confinement regions around the shock for relativistic electrons required field strengths of the order of $100$~$\mu$G, rather than just tens that could be expected based on compression of interstellar magnetic fields alone \citep[e.g.~][]{2003VinkLaming,2005Voelketal}. The non-resonant instability developed by \citet{2004Bell} resolved this. However, care should be taken by translating a magnetic field strength inferred from observations of TeV electrons to the magnetic field strength confining the PeV cosmic rays.

On scales smaller than the gyroradius effective growh of magnetic fluctuations is generated because of a plasma current. The current is a reaction to the cosmic-ray current, which requires balancing in the local plasma. Magnetic field fluctuations perpendicular to the current result in a $j \times B$ force that, for a given polarisation, focusses the cosmic rays within the growing magnetic loops, and pushes the plasma outwards. With a frozen-in magnetic field, the stretching of the field lines results in growth. In the approximation that the larmor radius of the cosmic rays is much larger than the fluctuations, the cosmic ray current is effectively undisturbed and this can result in significant magnetic field amplification. Simulations of this instability in the non-linear regime indicate that this indeed is the case \citep{2004Bell}. An estimate for the saturation value of the magnetic field results in this value being proportional to the shock velocity to the power $3/2$. Therefore, especially for young remnants where the shock velocities are high, magnetic field growth to orders of mG seems to be feasible.

The scales of the fluctuations triggered by this non-resonant instability are smaller than the gyro-radii of the driving cosmic rays. Scattering, being most efficient for cosmic rays having a larmor radius of the same order as the magnetic fluctuations, can become very efficient. Cosmic rays with low energies can therefore be confined near the shock region. However, in order to confine the high-energy end of the cosmic rays, magnetic fluctuations need to grow on a scale larger than the gyroradius of the dominant cosmic ray component. 

In a recent work, we describe the growth of magnetic fields on scales longer than the gyroradius, due to the influence of the stress tensor arising from gradients in the perturbed cosmic ray current \citep{2011SchureBell}. Small-scale turbulence, resulting from the non-resonant instability on small scales, induces an effective diffusivity that couples the right- and the left- hand polarisation through its action on the plasma.

From the dispersion relation in \citet{2011SchureBell}, the influence of contributions from the stress tensor and collisions can be evaluated. The full dispersion relation, taking into account the stress tensor and collisions, looks like:

\begin{eqnarray}
\label{eq:dispf2}
\omega^2=&&
\mp\Omega^2\left(i\omega-\frac{k^2 c^2}{5(3 \nu\mp i\omega_g)}\right)
\\\nonumber &&
\left/
\left(\nu - i\omega \mp i\omega_g+\frac{k^2 c^2}{5(3 \nu\mp i\omega_g)}\right)\right..
\end{eqnarray}
The upper signs correspond to the left-handed polarisation (similar to the polarisation of a gyrating proton) and the lower signs to the right-hand polarisation (the one that is unstable to the small-scale nonresonant Bell instability). 
The magnetic fluctuations grow when the imaginary part of the square root of the above equation is positive. $\omega$ is the complex frequency, $k$ the wavenumber, $c$ speed of light, $\nu$ the effective scattering frequency, $\omega_g$ the gyrofrequency, and $\Omega$ the growth rate of the non-resonant Bell instability that contains information on the driving current: $\Omega=\sqrt{k j_0 B/(\rho c)}$, with $B$ the unperturbed magnetic field strength parallel to $k$, $j_0$ the return current, and $\rho$ the plasma mass density.

The terms including factors of $k^2$ result from the inclusion of the stress tensor. Effects from the small-scale non-resonant instability are included through $\nu$, the effective scattering frequency. Scattering on small scales is expected to arise earlier than on long scales, since the growth rate increases rapidly for $kr_g \gg 1$.
Bohm diffusion is the regime where the effective scattering frequency is of the same order as the gyrofrequency, i.e. $\nu \approx \omega_g$.

The method in \citet{2011SchureBell} recovers the 2004 growth rate for the right polarisation in the regime where $kc \gg \omega_g$.
The resonant instability is not directly captured in the mono-energetic approach. When the collisionality is zero, $\nu=0$, the only growing mode on small scales is the right-hand one, whereas on long scales only the left-hand mode is unstable. When $\nu \ne 0$ both modes are unstable in the long-wavelength regime, although the left-hand mode continues to dominate until $\nu/\omega_g > 1/\sqrt{3}$. 

 \citet{2011Bykovetal} also investigated an instability on the large scales that takes into account the effect of mean field dynamo theory. Averaging of the magnetic fluctuations on small scales, arising from the 2004 Bell instability, generates a ponderomotive force that acts to amplify the field on large scales. They found rapid growth, and their analysis also indicates that the left-hand mode is the dominant growing mode in the long-wavelength regime.
A third option of growth on larger scales is that of growing small scale fluctuations. The filaments resulting from the small-scale nonresonant instability can merge and grow in scales \citep{2011RevilleBell}. 

\section{Typical growth times}

\begin{figure}
\begin{center}
\includegraphics[width=0.5\textwidth]{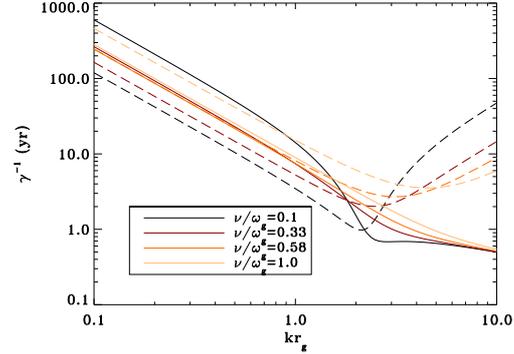}
\caption{Growth time in years as a function of wavenumber. The solid (dashed) lines represent the right- (left-) hand polarisation. Values are given for different efficiencies of the small-scale scattering rate. On long scales the left-hand mode has the quickest growth, whereas on scales smaller than resonance, the right-hand mode dominantes. }
%19/07/11:Phys/Instability/plotgrowthrates.pro
\label{fig:growthtimesB3}
\end{center}
\end{figure}

The driving current is a result of the streaming of cosmic rays, being proportional to the shock velocity: $j_0=n_{\rm cr} q v_{\rm s}$, with $n_{\rm cr}$ the cosmic ray number density, $q$ the unit charge, and $v_s$ the shock velocity. Typical values for growth rates are summarized in Table~\ref{tab:growthrate} and plotted in Figure~\ref{fig:growthtimesB3}. A current density of $j_0=10^{-8}$~stA~cm$^{-2}$  is assumed at the shock, corresponding to $n_{cr} \approx 10^{-5}$~cm$^{-3}$ and $v_s=10^9$~cm~s$^{-1}$. 

If $n_{\rm cr}$ is constant, the current scales with $\left(\frac{E}{TeV}\right)^{-1}\left(\frac{v_s}{10^9\;{\rm cm\;s}^{-1}}\right)$, given a powerlaw spectrum with a slope of $p=2$, and a specific scattering efficiency. The particle energy is that of the driving cosmic rays, which is increasing with distance upstream of the shock when diffusion propagates proportional to the particle energy. Bohm diffusion can be considered to be the minimal diffusion rate. Dividing this by the shock velocity gives a lengthscale to which particles can diffuse upstream, given by $L=D/v_s=(r_g c)/(3 v_s)$. 
The corresponding time scale can be found by dividing again over the shock velocity, to determine when the shock has caught up with the location where the instability is triggered, giving a lower limit of $\tau=c r_g/(3 v_s^2)=0.35  \left(\frac{E}{TeV}\right)\left(\frac{B}{3 \mu G}\right)^{-1}\left(\frac{v_s}{10^9\;{\rm cm\;s}^{-1}}\right)^{-2}$~yr. The growth time scales with $\left(\frac{E}{TeV}\right)\left(\frac{B}{3\mu G}\right)^{-1}\left(\frac{v_s}{10^9\;{\rm cm\;s}^{-1}}\right)^{-1/2}$, for $n_{cr}=10^{-5}$. The number of e-foldings for growth, $\gamma \tau$, can be seen to scale with $v_s^{-3/2}$, leaving only the shock velocity as a free parameter to increase the linear growth effectively.  

\begin{table*}
\caption{Linear growth times for the right (and left between brackets) hand modes, given for a particle energy of $1$~TeV, magnetic field strength $B=3\,\mu$G, shock velocity $v_s=10^9$cm~s$^{-1}$, and current $j_0=10^{-8}$~stA~cm$^{-2}$. It can be derived for other values by using the fact that it scales with $\left(\frac{E}{TeV}\right)\left(\frac{B}{3\mu G}\right)^{-1}\left(\frac{v_s}{10^9\;{\rm cm\;s}^{-1}}\right)^{-1/2}$ if $n_{cr}$ is constant in time, or with $\left(\frac{E}{TeV}\right)\left(\frac{B}{3\mu G}\right)^{-1}\left(\frac{v_s}{10^9\;{\rm cm\;s}^{-1}}\right)^{-3/2} \left(\frac{\eta}{0.014}\right)^{-1/2}$ if $n_{cr}$ is a fraction of the shock kinetic energy. }
\label{tab:growthrate}
\begin{center}
\begin{tabular}{ l l l l l l l }
\hline\hline
\\
$\nu/\omega_g$ & $\gamma^{-1}$ (yr) & $\gamma^{-1}$ (yr) & $\gamma^{-1}$ (yr) & $\gamma^{-1}$ (yr)& $\gamma^{-1}$ (yr)& $\gamma^{-1}$ (yr) \\
 & $(k r_g=0.1)$ & $(k r_g=0.5)$ & $(k r_g=0.8)$ & ($k r_g =1$) & ($k r_g =5$) & ($k r_g =10$)      \\
\hline
\\
1/10 & 600 (119) & 50.2 (10.5) & 22.1 (5.00) & 4.47 (1.10) & 0.651 (13.1) & 0.498 (47.4) \\
1/3 & 264 (164) & 23.0 (14.7)  & 10.8 (7.31) & 2.34 (1.67) & 0.700 (4.62) & 0.503 (14.6)  \\
$1/\sqrt{3}$ & 245 (246) & 21.7 (22.2) & 10.6 (11.2) & 2.35 (2.58) & 0.776 (3.53) & 0.513 (8.92) \\
1 & 283 (462) & 25.3 (41.6) & 12.5 (20.9) & 8.95 (15.2) & 0.949 (3.58) & 0.542 (6.01) \\
\hline 
\end{tabular}
\end{center}
\end{table*}

Alternatively, it can be considered that $n_{cr}$, and therefore $j_0$, scales with the kinetic energy of the shock, such that a fixed proportion of $\rho v_s^2$ is converted into cosmic rays. For a mono-energetic approach, this translates to $n_{cr}E_{cr}=\eta \frac{1}{2}\rho v_s^2$ such that the current density becomes $j_0=n_{cr}qv_s = \eta q \frac{1}{2}\rho v_s^3/E_{cr}$ and thus scales with $\left(\frac{E}{TeV}\right)^{-1}\left(\frac{v_s}{10^9\;{\rm cm\;s}^{-1}}\right)^3$. The growth time in that case scales as $\gamma^{-1} \propto \left(\frac{E}{TeV}\right)\left(\frac{B}{3 \mu G}\right)^{-1}\left(\frac{v_s}{10^9\;{\rm cm\;s}^{-1}}\right)^{-3/2}\left(\frac{\eta}{0.014}\right)^{-1/2}$ and the number of e-foldings is proportional to $\gamma \tau \propto v_s^{-1/2}$. Of course in these equations $\left(\frac{E}{TeV}\right)\left(\frac{B}{3 \mu G}\right)^{-1}$ can be replaced by $\left(\frac{r_g}{1.11 \times 10^{15} {\rm cm}}\right)$, with $r_g$ the gyroradius. The numbers in Table~\ref{tab:growthrate} correspond to a value of $\eta=0.014$. 
It is expected that the fraction of kinetic energy going into cosmic rays can be significantly higher than this, resulting in shorter growth times by a factor of the square root of this efficiency. For an efficiency of 10\% this would result in reduced growth times compared to those in Table~\ref{tab:growthrate} by a factor 2.7.

\section{Discussion and conclusion}

Acceleration of cosmic rays in supernova remnants requires the growth of magnetic fluctuations in order to confine cosmic rays to the shock region. For the highest-energy cosmic rays the non-resonant regime on scales larger than the gyroradius of the dominant cosmic ray component is of importance in confining cosmic rays as they reach higher energies. A non-resonant instability is discussed that is driven by the return current in the plasma, resulting from a reaction to the cosmic ray current. The stress-tensor is crucial in generating an instability in the long-wavelength regime. Scattering on small scales aids in mediating the instability between the two polarisations. The number of e-folding times for the instability at a given scattering efficiency is shown to depend only on the shock velocity, and typical growth times are given.

\begin{acknowledgements}
This research was supported by the UK Science Technology and Facilities Council grant ST/H001948/1.
\end{acknowledgements}

\bibliography{../../adssample}

\label{lastpage}
\end{document}